# FEMALE STUDENTS IN COMPUTER SCIENCE EDUCATION: UNDERSTANDING STEREOTYPES, NEGATIVE IMPACTS, AND POSITIVE MOTIVATION

Bernadette Spieler, University of Hildesheim[1], Hildesheim, Germany, *bernadette.spieler@uni-hildesheim.de*
*Libora Oates-Indruchovà, University of Graz, Graz, Austria, libora.oates-indruchova@uni-graz.at*
*Wolfgang Slany, Graz University of Technology, Graz, Austria, wolfgang.slany@tugraz.at*

*Although female students engage in coding courses, only a small percentage of them plan to pursue computer science (CS) as a major when choosing a career path. Gender differences in interests, sense-of belonging, self-efficacy, and engagement in CS are already present at an early age. This article presents an overview of gender stereotypes in CS and summarizes negative impressions female students between 12 and 15 experience during CS classes, as well as influences that may be preventing girls from taking an interest in CS. The study herein draws on a systematic review of 28 peer-reviewed articles published since 2006. The findings of the review point to the existence of the stereotypical image of a helpless, uninterested, and unhappy "Girl in Computer Science". It may be even more troubling a construct than that of the geeky, nerdy male counterpart, as it is rooted in the notion that women are technologically inept and ill-suited for CS careers. Thus, girls think they must be naturally hyper-intelligent in order to pursue studies in CS, as opposed to motivated, interested, and focused to succeed in those fields. Second, based on the review, suggestions for inclusive CS education were summarized. The authors argue that in order to make CS more inclusive for girls, cultural implications, as well as stereotypization in CS classrooms and CS education, need to be recognized as harmful. These stereotypes and cultural ideas should be eliminated by empowering female students through direct encouragement, mentoring programs, or girls-only initiatives.*

## 1 INTRODUCTION

Our future will be digital, and the next generation of jobs will be characterized by an increased demand for people with computational and problem-solving skills (Box, 2018; European Statistics Eurostat, 2019a). Knowledge in CS or even a university degree in CS is thus a lucrative educational prospect. As a result, the importance of learning about digital technologies for young people has been heavily discussed in the media and by governments (Committee on European Computing, 2016; European Commission, 2016). The underrepresentation of women in the computing industry is an important dimension in discussions, as is the underrepresentation of women in computer science (European Statistic Eurostat, 2018b).

In 2006, a new concept by Wing (2006) focused on reinforcing Computational Thinking (CT) skills among students and promoted coding as a mandatory skill that opens a new way of thinking by introducing problem solving and critical thinking tasks in the classroom. Her findings have been incorporated into the US curriculum of many federal states of the US (Kahn, 2017), into K-12 movements (Mannila et al., 2014), and launched by Barack Obama in 2016 during the "Computer Science For All" initiative (Ladner & Israel, 2016). Also in Europe, teaching CT-skills is important. Many countries, including Austria and Germany, organize an annual Bebras contest that usually takes place in November[*]. The aim of this contest is to help students of different age groups to develop CT-skills, to boost interest in CS, to think about tricky tasks related to CS and to disseminate basic concepts of CS without the actual use of a computer. In addition, non-profit organizations like Code.org have supported different projects since 2013 which foster CS education, for example, with age-appropriate videos, games, and courses. Code.org also promotes the "Hour of Code" event, which provides one-hour tutorials in many different languages[†]. Promoting diversity and empowering girls plays a high priority in their agenda as well. For example, the curriculum materials are designed to recruit, engage, and retain women and people of color and to share inspiring videos and role models. As a result of this targeted campaign, statistics of their courses show a high amount of female participants. Other institutions bolster girls in CS in particular via girls-only initiatives. For example, GirlsWhoCode[‡] from the US, or Codefirst:Girls[§] from the UK offer services only available to girls, giving them a place to explore coding without feeling pressured to compare themselves with male counterparts who may already have more experience or familiarity with CS.

Two more trends emerged around 2006, which contributed to the success of the ideas of teaching CT skills to all students. First, a number of block-based visual oriented programming tools were introduced which should help novice programmers and young learners in their first programming steps. The most popular visual coding environment is Scratch[**], introduced in 2007 by the MIT Lifelong Kindergarten Group (Resnick et al., 2009). Unlike traditional programming languages, which require code statements and complex syntax rules, Scratch uses graphical programming blocks that automatically snap together like Lego blocks when they make syntactical sense (Ford, 2009). Second, a widespread use of smartphones can change how learning takes place in many disciplines and contexts. Mobile devices are more frequently used than computers or tablets and from the age of 12 almost all are online (Bitkom, 2019). They are already a part of our culture and for most adolescents the smartphone performs several functions of their daily lives. As a result, a new generation of young digital natives emerged.

---
[1] with institutional support of Graz University of Technology

Girls and women have caught up with boys and men as users of this technology but they are not its producers (Siobhan, 2018). If women are absent from this field, they will not participate in the shaping of the world of technology and in the decision-making on technological research agendas. Recent statistics from Europe and all over the world indicate that women's representation in Information and Communication Technology (ICT) professions and related university programs have not improved and women's employment in these occupations increased at a slower rate than men's employment in the past decade (Eurostat Statistics, 2017-2019). Jobs and university programs in ICT are clearly male-dominated, as illustrated in selected statistical data by European Statistics Eurostat from 2015-2017. For example, in 2017, 83% of the people working as ICT specialists were male (Eurostat Statistics, 2019a; Eurostat Statistics 2018c). Further data gathered by the European Statistics Eurostat (2017b, 2018b) confirms a low percentage of female students in studies related to ICT. In 2016 in Austria, the percentage was only about 14%, which is even lower than the EU average (17%), but higher than the share of women in ICT studies in Luxembourg (10%), Belgium (8%) and in the Netherlands (6%). In the UK, the numbers fell from 15% of CS graduates in 2016/2017 down from 16% in 2015/2016 (Wise Campaign Report, 2017). However, the gender gap in ICT is not a worldwide phenomenon and does not exist to such a degree in Eastern Europe, for example in Romania, Bulgaria, or Latvia (Eurostat Statistics, 2019a). The countries of the former Eastern Bloc have a strong tradition of mathematics and sciences because of educational ideals reinforced in communist regimes. The regimes prioritized hard sciences over social sciences and humanities and subjected them to lesser surveillance and censorship (Sanders, 2005; Ciobanu, 2018; Oates-Indruchovà, 2020). As a result, female students have more role models that may provide a partial corrective to cultural stereotypes. Outside of Europe, the underrepresentation of women in ICT is a topical issue as well. In the US, women hold just 11% of executive positions at Silicon Valley companies (Bell & White, 2014) and white men dominate college CS departments (National Center of Education Statistics, 2019). In 2017 the numbers in Australia show that although 46.2% of women make up the workforce, only 28-31% of workers in the roles of technology are women (Charleston, 2017). Also alarming is the digital gap in developing countries, as study of the Nigerian town of Kafanchan suggests (Buskens & Webb, 2014). According to statistics by Dalberg Design (2015) men in Africa are twice as likely to have access to the Internet as women; this is especially true for regions like sub-Saharan Africa (45% of the women have no internet access). Furthermore, in Israel, 61% of the students in CS classes are female. However, a closer look shows that countries that have the most female college graduates in STEM and in CS in particular were also some of the least gender-equal countries (Stoet & Geary, 2018). The article by Ripley (2017) states "[A] boy doesn't need to study hard to have a good job. But a girl needs to work hard to get a respectable job." In addition, the overall proportion of the female workforce relative to the overall labor force is very small across Arabic countries (The World Bank, 2017), see Saudi Arabia (16.19%), or Yemen (7.88%). Statistics from Asia show mixed numbers. On the one hand, in Malaysia (60% female students in CS programs) or at Chang Gung University in Taiwan and Mahidol University in Thailand (both 48% female students in CS programs), no gender gap in IT is present (Ong, 2016). On the other hand, at the Hiroshima University in Japan, only 4% of the students are female. In India, the problem women face is a more general barrier to work than a gender gap in a specific sector (Oliver, 2017).

To conclude, a comparison of "women and technology" across cultures on one dimension raises more open questions than it brings solutions. It seems that countries that empower women empower them more to choose the career they are interested in the most. However, it is well recognized by many authors (e.g., suggested by Bers et al., 2014; Gabay-Egozi et al., 2015; Unfried et al., 2015; Zagami et al., 2015; Beyer, 2016; Khan & Luxton-Reilly, 2016; Ko & Davis, 2017) that gender gaps in CS do not start at industry or university level but between the ages of 12 to 15, or even earlier in childhood. The first comprehensive analyses on young girls' motivation for CS was the research by Margolis and Fisher (2002) examining impacts at college and high school level which contribute to the gender gap in CS. On the one hand, their research and further studies by Carter (2006) and Sadler et al. (2012) documented girls' low experience levels towards computer sciences, hence, contact with technology at an early age could perhaps change the attitude of women before stereotyping discourages them (Carter, 2006; Sadler et al., 2012). On the other hand, researchers stated that female students either do not feel motivated to learn about technology nor they are encouraged by friends or their family (Unfried et al., 2015; Master et al., 2016).

The present review summarizes differences in girls' motivation for CS at the age between 12 and 15 years with the goal to illustrate the socially constructed stereotype of a girl in computer science and negative implications. This set the background for more practical exploration projects, namely, for the "RemoteMentor" research project, the "Girls Coding Week" and the "Code'n'Stitch" project, conducted by the authors and described in the last section of this paper. This article is structured as follows: Section 2 describes the methodology used for the systematic review of research on stereotypes, negative impacts, and positive motivation in CS, while Section 3 presents the results of the articles selected. In Figure 1, the authors summarized the gender differences in perception of motivation for CS. By drawing on the defined research questions, the individual parts of the figure are described in greater detail by showing how stereotypes and negative implications in CS are (re-)constructed, explaining differences in CS classrooms, and pointing out inequity in CS-education. The discussion section, Section 4, shows empowerment strategies to create inclusive CS activities for all genders. Strategies are illustrated in Section 4 and Section 5 concludes the paper. Finally, the last section, Section 6, presents promising projects by the authors to engage female students which were influenced by this review.

## 2　　SCOPE AND PURPOSE OF THE SYSTEMATIC REVIEW

For the present systematic review, research questions have been elaborated and the search strategy as well as the criteria of the articles selected for the review have been described as proposed by Kitchenham and Charters (2007). Finally, the extracted data (the articles) is presented.

## 2.1 Research Questions

To help us to formulate the research questions and to focus on the general concept, the PICOC (Population, Intervention, Comparison, Outcomes, Context) method has been used (Petticrew & Roberts, 2006), see Table *1*.

"Population" has been defined with "students between 12 to 15 years old" and "students in CS majors". The reason for selecting the first was that secondary school and the first years of high school (girls between 12 to 15 years old) are educational phases during which young people decide their future career orientations, develop a more realistic picture of their future jobs, and assess their career-relevant abilities (Appianing & Eck, 2015). Because high school is where many female students begin to lose interest in scientific fields or self-confidence in relation to their male peers, we focused a great deal of our research on students at this level. The students at university level themselves play an important role because they are shaping the CS culture and environment from the inside (Frieze & Quesenberry, 2015). We begin with reviewing literature on CS in secondary education, followed by experiences of female students actually studying in a CS-related field. This will help us to identify differences in motivation between girls and boys that participate in creating preconceptions and assumptions about CS during education experience and everyday life.

Therefore, the research questions for this review were:

**RQ1:** Which cultural and social aspects in CS have been identified that influence girls' motivation for CS?

**RQ2:** What preconceptions and differences in behavior of girls towards CS have been identified?

**RQ3**: What differences were identified in unequal treatment of girls and boys in CS classrooms?

**RQ4**: Did articles detect differences in coding and game development between the sexes?

Research Question 4 has been added because the first author of this paper has software engineering as her major and the programming projects she administered are described in Section 6. By answering these four research questions, the aspects which strengthen the construction of stereotypes and negative implications at this age (secondary school and high school) should become apparent.

## 2.2 Search Strategy and Process

The search strategy considered the "Intervention and Outcomes" criteria of PICOC. Keywords in searches included:

- ("Girls" OR "Gender") AND "Computer Science Education"
- "Stereotypes" AND "Computer Science Education"
- ("Coding" OR "Programming") AND ("Girls" OR "Gender")

To find evidence for RQ1, the following keywords related to cultural and social influences have been added:

- ("Role Models" OR "Mentors") AND "Computer Science Education"
- ("Girls" OR "Gender") AND ("Coding" OR "Games" OR "Game Design")

The search was performed using online databases that include references for scientific articles and journals and conference proceedings: ACM, IEEE Xplore, SAGE Journals, Wiley Online Library, APA PsycNET, and Springer.

## 2.3 Study Selection

We have defined a set of inclusive and exclusive selection criteria to identify the most appropriate articles for the systematic review that helped us in answering the proposed research questions. Articles were included, which focused on girls between 12 to 15 years in CS classrooms or female students in CS university degree programs. Furthermore, we focused on empirical literature gathered since 2006 with the aim of providing a compact summary of findings since the mainstream introduction of coding in different school levels. The corpus of articles covers CS research carried out in the US, Australia, and Western Europe, countries that face similar problems in attracting female students to CS (see Section 1).

Studies that fulfilled either of the following criteria were excluded:

- Focus on STEM (Science, Technology, Engineering and Mathematics) in general. These articles were taken into account if they made a clear distinction between CS results and other STEM subjects.
- Present reviews or similar systematic mappings.

For example, existing literature reviews cover summaries on retention of women in CS (Pantic & Midura, 2019), on women in computer-related majors (Allen, Scheckler & Darlington, 2007), or consider only girls' interest in computing to diverge from boys' (Cohoon & Asprey, 2006). The aim of our research is to present a comprehensive picture of the issues emerging in CS in girls between 12 to 15 years old.

- Abstract only, posters, works in progress, and panel papers.
- Articles that refer to a specific discipline of CS (besides coding), for example AI, smart homes, robots, chat bots etc.

## 2.4 Extracted Data and Sample of Articles

The results in Table 2 included the following information about the included articles: (a) Author(s), (b) Publication title, (c) Publication date, and (d) Database in which the articles were found. The last column shows the allocation to the corresponding research questions defined for this review. We obtained 318 studies from the consulted databases. After applying the inclusion and exclusion criteria, we had a sample of 28 articles that were selected for this review.

These were supplemented with recent studies conducted at Harvard, by Microsoft, and university presses such as MIT Press, Oxford Academic, Cambridge Journal Education, and American Educational Research Association. Furthermore, Carnegie Mellon University had a great success in the last several years in recruiting more female students for CS and the results published in the form of a book have also been considered for this review (Frieze & Quesenberry, 2015). Furthermore, a white paper by Google Inc. (2018) examined females' playing behavior as part of the "Change the Game" campaign. These findings were an important contribution to investigate differences in playing behaviors on smartphones.

## 3 RESULTS

The results of the review were summarized in Figure 1. This figure represents the authors' summary of gender differences in perception of motivation supplementing with attributes and stereotypes associated with young boys and girls in CS. Stereotypes in general influence people and produce misrepresentations (Matlin, 1999). They describe specific behaviors, attitudes, and capabilities purportedly associated with, for example, a certain class, race, gender, or profession. Since CS is stereotypically associated with the masculine role in the still prevalent gender binary, female teenagers are less likely to have a sense of belonging vis-à-vis CS, may feel less interested, and engaged in these classes (Master et al., 2016). A method called "dramatization" of gender foregrounds the binary divisions by gender in order to develop sensitivity and awareness (Engler & Wieland, 1995). Dramatization of gender is a standard for a reflective gender practice by promoting specific perception of individual interests and abilities. Naturally, the "dramatization" method has its limitations as it works with averages rather than individualities, differences rather than similarities, and it does not account for the plurality of genders. As such, it runs the risk of contributing to the strengthening of the very stereotypes it aims to expose and to the reinforcing of the gender binary. We used the method here in the sense of Spaak's "strategic essentialism" to draw attention to the lasting effectiveness of socially constructed categories in CS (Spivak, 1990). The dramatization will serve as a springboard for the typology of research that critically engages with the stereotypes and negative implications in CS that we present below. The following chapter provides details to Figure 1 by answering the Research Questions in 2.1, and describing the selected articles for this review in a descriptive analysis.

## 3.1 Cultural and Social Influences, and Family Support

In general stereotypes disempower women in their field at all levels (teenage girls, undergraduate female students, and even successful female scientists; Young et al., 2013; Galdi et al., 2014). For example, the geeky, nerdy, isolated, fanatical computer expert who represents the stereotype of CS is certainly not something most young women strive to become (Cheryan et al., 2013, Lewis et al., 2016). The research of Lewis et al. (2016) provide evidence that the technology sector needs people with diverse skills, those who can design, develop, analyze, and manage information technology, rather than people with a narrow skills scope, such as only technology development or only programming. This section refers to RQ 1 and describes the reviewed articles that contribute to the formation of stereotypes through parental influences, the absence of female role models and mentors and social influences by games or media.

A survey of curricular choices of high school students by Gabay-Egozi et al. (2015) lists the reasons for gender-typical educational choices. Their results show the negative effects of parental preconceptions and influences, e.g., when families exclude computing or engineering as a possible career path for their daughters. While making educational choices, male and female students may follow these perceptions of an appropriate choice for their gender rather than their interests. The study points out that those teenage girls are less likely to choose a CS-related path unless they were encouraged by their parents, by teachers, or by their peers. Self-doubt in female students towards STEM disciplines occurs if they feel intimidated because they need more help especially in CS subjects, thus thinking it must be attributed to their deficient ability. If the profession does not fit the traditional gender model, one is not as likely to pursue or feel discriminated against by someone who does. To be socially connected and respected is a strong initial motivator; it can "create a sense of belonging that can reinforce students' self-efficacy and connections to community that support student perceptions of their ability within the field" (Veilleux et al., 2013, p. 64). According to that study, it is not important whether the father or the sister has a technical profession; rather, it is more about a general and supportive attitude of relatives or other important people in a girl's life that encourages her to follow her own path.

Lewis, Anderson, and Yasuhara (2016) elaborate how students assess their fit with CS. In their previous work the authors identified five factors that shape students' decisions for choosing or not choosing a major in CS (Lewis, Anderson & Yasuhara, 2011): 1) their ability as related to CS, 2) their fit between their identity and CS, 3) their enjoyment of CS, 4) the utility of CS, and 5) the opportunity cost associated with majoring in CS. In 31 semi-structured interviews, they asked students about their experiences in CS courses, academic interests, and preconceptions of CS. Their results showed that students evaluate their "fitting in" by measuring their self. This is done by focusing on characteristics associated with their perceptions on computer scientists (described as singularly focused, asocial, competitive, and male) as a result of their beliefs about stereotypes (shaped by observations in their CS courses and the experiences of their peers).

Stanko and Zhirosh (2017) examined the role of the family and choosing careers in IT or CS. The terms Information Technology (IT) and Computer Science (CS) are not used synonymously (KING University Online, 2017). While IT careers mainly deal with the installation, maintenance and improvement of computer systems or the operation of networks and databases, CS is focused on mathematical algorithms or design and development tasks. An IT career does not necessarily require a CS degree. Stanko and Zhirosh (2017) conducted for their study eight focus group discussions with 46 parents of boys and girls. Questions were asked about the development of their children's interests, their decision to work in CS or IT, and the role of parents in their children's decisions. Those parents who themselves work in IT taught their children directly. Those parents who do not work in these areas took on a supporting role if their children were already interested in CS-related topics. However, the authors found that family support is an important factor in the development of boys' and girls' interests and readiness in CS in general.

For a successful professional development, female mentors and appropriate role models are two more key elements that further strengthen girls' self-efficacy and interest (Stout & Camp, 2014).

### 3.1.1 Exposure to female role models.

Lockwood (2006) presents two studies that examined the importance of female role models for career choices in first-year female students. The results showed that 1) women prefer female career role models to male ones, and 2) that they derive special benefits from gender-matched role models, by naming their achievements. It is important that women see female role models who have succeeded and who promote positive beliefs regarding women's abilities, which demonstrates that this job field can suit them as well. Since the technology sector is very male-dominated, most role models are also male. However, the study emphasized that women who had to overcome gender stereotypes or are successful in traditionally male-dominated fields are most effective as role models.

Cheryan et al. (2011) argue based on two experiments with female non-CS majors that role models who were perceived as very stereotypical can have a negative influence on women's choices. They give as examples role models who prefer stereotypical games or movies (e.g., Star Wars, Star Trek, etc.), have a very stereotypical appearance (e.g., the geeky female computer scientist who wears glasses), or a role model who is a supernatural genius in her field (e.g., a girl who had already started programming at the age of 11). Women who confirm the male stereotype have a negative effect on women's beliefs about their ability to be successful in STEM in general. If role models are presented as somehow supernatural, it implies to girls that only outliers and geniuses would succeed in a field that is not typical for their gender.

A study conducted by Young et al. (2013) provides further evidence of the importance of meaningful contact with female role models. They mostly refer to university professors but also to female role models at every stage, e.g., female teachers at the secondary level in mathematics or science, female students, faculty members in higher education, and women working in STEM fields. These "ordinary" role models are associated with pro-science career aspirations and attitudes because it is easier to identify with them, and as a result, the subject is associated more with women. They help reduce or even invert the implicit stereotype. The authors concluded that female professors were viewed as more positive role models than male professors were.

Dele-Ajayi et al. (2018) built a gaming environment to generate interest and engagement among young people. To create awareness of professional career roles in CS, a sorting game was introduced at the beginning. This allowed participants to assign job roles to specific characters. The results showed that boys tend to assign more men to technology related job positions (e.g., programmer) and girls used a mix of male and female characters. In addition, girls were more likely to create games with female characters than boys were. The results show that young people already have a particular perception of the digital games sector and show gender-specific preferences.

Semmens, Piech, and Fried (2015) asked 162 high school girls ages 15 to 17 in a summer program, to list attributes that describe themselves and those of a computer scientist both before and after an eight-week CS program. Those attributes mentioned at the end of the course were more positive and less stereotypical. During their course program the girls had opportunities to engage with teaching assistants and guest speakers in CS professions which provide them with a more realistic picture of people working in that field. In addition, these participants saw themselves after the course as more similar to a computer scientist, showing how important it is to introduce girls to realistic role models. These earlier findings find an echo in a recent ISAC global survey report (Wisniewski, 2017): the absence of female role models is currently one of the top five reasons why women are underrepresented in technology (42% women mention this in the findings of this ISAC study).

### 3.1.2 Supporting mentoring (-programs).

A number of mentoring programs have appeared in the last years[††]. Mentors not only act as role models but additionally provide guidance, for example by increasing one's self-confidence during a task completion, or during the first semesters of university study. A study by Clarke-Midura et al. (2016) evaluated programs designed to encourage female high school students to volunteer for mentoring programs to help girls from middle school in CS (with the goal of empowering both age groups). This near-peer mentoring provided a positive experience for both groups in terms of interest and self-efficacy.

Two studies by Ko and Davis (2017) investigated if mentoring enhances students' interest, beliefs, and engagement level in CS. The results showed that students who had mentoring relationships with different kinds of people, including friends, parents, siblings, cousins, teachers, and even neighbors, were influenced positively. For example, they were more interested in computing (e.g., incorporating computing into their identities), had a more positive beliefs about people working in CS (e.g., the associated attributes were "creative", "patient", "intelligent", or "hard-working"), and finally, felt more engaged in coding opportunities (e.g., encountering more programming languages).

The report by The Association for Women in Science (2017) clearly identifies mentoring as being

> "[A]mong the most effective ways to bring about change, because it reaches across institutions, fields, and even generations. Mentoring unifies women, validates their experience, provides inside information about the system's workings, trains groups of individuals to challenge the system, and provides the basis for generations of outsiders to both enter the system and, as insiders, demand needed change."
> (The Association for Women in Science, 2017)

In male-dominated fields, female students need encouragement. However, the study did not indicate if mentors must be women or if it is only important to have somebody who believes in them and encourages them.

### 3.1.3   The influence of movies and games.

Numerous studies show that the movie and gaming industries contribute to the reinforcement of stereotypes (Kimmel, 2004; Barker & Aspray, 2006; Beekhuyzen & Dorries, 2008). At the beginning of the 20th century, computers and computer programmers were seen as something negative and strongly associated with women and low-paid secretarial occupations (Brewer, 2017). This image has changed dramatically over the last several decades. Researchers have tried to explain the huge decline from approximately 50% to 37% of women who are working in technology in 1986 to a mere 17% today (Ashcraft et al., 2016; National Center for Women and Information Technology, 2016). Possible reasons for this phenomenon lead to the influence of pop culture and targeted marketing (Brewer, 2017). For instance, games in the 1980s established a stereotype of the "computer programmer" (e.g., "WarGames") and the development of video game consoles and games targeted mostly at young boys. Video, computer, and mobile games are not only entertaining and fun. The study by Davies et al. (2014) show that there exists a correlation between students who do not play video or computer games and those who describe their computer skills as insufficient or do not spend much time on technical devices. At the same time, jobs in technology became more lucrative and "cool", which appealed to boys and men hoping to secure high paying jobs, prestige, and technology.

Jenson, de Castell, and Fisher (2007) present findings from a three-year, federally funded Canadian research project that promoted a console-playing club. A number of 44 boys and 60 girls were interviewed and completed a questionnaire about their gaming behavior. Findings at the beginning of the study showed that the girls were not used to playing console games and most were just 'watching' their brothers or other male relatives while playing. When playing console games, they needed more instructions whereas the boys already knew how to play or "learned by doing". Over the span of three years, girls helped each other more or/and become more competitive and self-effacing. They started to compare game scores and showed interest in competitive play. In boys' groups, playing was strongly connected to their identity (i.e., winner or loser) – this continued throughout the study. Additionally, at the end of the project all boys reported playing with other boys and all of the girls reported playing with boys, and only infrequently, with other girls. The author explained this with the previously stated fact that girls were used to playing with males. To conclude, gender differences were far less evident or no longer present at the end of the study. The authors argue that this is due to the support girls received, or that they had the right to choose between games (and were not restricted to those their brother or others wanted to play). The study supports the argument that engagements in which boys traditionally appear can intimidate girls. The authors suggest that researchers should focus less on game preferences between genders and more on what conditions girls tend to play under.

Oversexualization of women in games is one other contributing aspect to the reinforcement of the binary gender stereotypes in CS. A study by Lynch et al. (2016) shows that the trend peaked in 1995 and then declined, but that women still occupy secondary roles and are objectified more often for several reasons (e.g., marketing, lack of female game developers, or lack of game developers who are aware of gender issues). A closer look at the existing video games shows that nowadays, more and more strong women occupy the main role in video games. One example from 2017 is a redesign of the famous character Lara Croft[‡‡] in Tomb Raider, who became a much more realistic young woman in her most recent incarnations. Another examples are the hero Aloy[§§] from "Horizon Zero Dawn", an outsider with an interesting story, or "The Scythian"[***] from "Sword and Sorcery", a video game for mobile platforms (she demonstrates that women can also be heroes).

Nevertheless, video games with female protagonists are still in the minority. Google, as one of the biggest app store providers, collaborated with the gaming intelligence provider NewZoo (2017) to examine American female players' experiences and perceptions by performing a quantitative study published as a white paper (3,330 female participants aged 10-65; Google Inc., 2018). They examined that video games appeal more to a male audience and exclude female gamers, there are fewer successful games available for girls than there are for boys, most games do not appeal to a female audience, and girls are less motivated to become gamers; only 30% consider themselves to be gamers. A survey by Yee (2017) conducted with 1,266 gamers shows that 75% of the female gamers rated female protagonists as "very" or "extremely" important; more than three times as much as male gamers. However, male characters in games prevail and are not inclusive at all. For instance, Grand Theft Auto[†††] shows sexual harassment and killing of female characters, and female characters tend to be scantily clothed (e.g., Juliet Starling from Lollipop Chainsaw[‡‡‡]). The study done by Google concludes that game developers have to rethink their strategies that games are becoming more appealing to wider audiences, in order to target all cultures, genders, and interests. In addition, Google encourages developers to put more emphasis on the personalities, emotions, and backstories of characters in the games because there exists less motivation for girls to become gamers.

## 3.2   Preconceptions and Differences in Behavior in CS Classrooms

This section summarizes articles that provide evidence for RQ2. Reviews show that younger female students (12 years old) do not show significant disinterest in CS related topics, but that the majority of teenage girls decide against obligatory CS related courses during high school (Zagami et al., 2015). Especially in girls between 13 to 14 years, female participation or interest toward STEM subjects and CS in particular begins to decline (Beyer, 2016). Thus, many studies conclude that high school is the key to understanding the field-major segregation in higher education (Sadler et al., 2012; Mann & Diprete, 2013; Gabay-Egozi et al., 2015; Unfried et al., 2015). In the higher grades, male students express more interests towards physics and engineering, while girls are more likely to prefer biology (Gabay-Egozi et al., 2015), medicine, and health (Sadler et al., 2012). These preferences reflect typical gendered educational choices resulting from the gender socialization influences that were discussed above. Nevertheless, reasons for a low interest in a subject may not always be gendered, but can originate in, e.g., teachers' failure to attract or explain the subject well, or in the low curriculum allocation. This is the case in many European countries, including the authors' country, Austria, with regard to CS, where classroom time for these topics is limited (Ebner & Grandl, 2017). Both groups of problems can lead to mis- or preconceptions in CS.

The following section surveys the research on the differences in students' interests, experiences, confidence, self-efficacy, fun, and engagement level in more detail.

### 3.2.1 Interest in CS activities and CS prior experiences.

The study of 836 high school students by Carter (2006) shows that boys were more likely to have prior experiences with CS (40% boys, 27% girls), more prior knowledge about CS (26% of boys versus 17% girls were able to provide a description of the CS major). In the latter case the answers included: "programming", "networking", "advanced use", "repair", "how computers work", "computer stuff", "good understanding". Most students of both sexes, however, had "no idea" (around 650 answers = 80%). However, even if students had no idea of what computer scientists actually do or what programming is, 11% of them mentioned that they stay away from CS as a major because of programming. Other answers (occurring with a similar frequency in both sexes) included "sitting in front of the computer all day", and that they would prefer a more "people-oriented major". The most interesting finding was the rating of a combination of CS with another field of interest, such as business or medicine, as a positive influence for choosing CS as a major (rated as the number one positive influence for girls). However, computer games or previous experiences were also mentioned as positive influences (more by boys than girls). The study suggests improvements for CS education, e.g., to fix the image of CS (it is not only about programming), allow (female) students to gain CS experience in high school (to avoid misconceptions about that field), and make CS courses fun (make them more creative and relevant). Low or bad experiences lead, of course, to bad expectations: if young women are used to failing or think they will fail in a specific course, they will instead choose subjects that they know they are comparatively good in. Often, all it takes is one bad classroom experience to scare students away from a discipline.

A cohort study of more than 6,000 students by Sadler et al. (2012) showed that the interest during high school changes dramatically for female students whereas the percentage of males interested in STEM remains the same. Particularly low female interest rates were found in astronomy and CS. The percentage of female students reporting an interest in STEM fell from 12.1% at the beginning of high school to nearly half of that (7.6%) by the end of high school. By comparison, three quarters of the male students said that at the beginning of high school that they were already interested in STEM and that percentage remained more or less unchanged at the end. This gender disparity applied, particularly, to engineering subjects. The initial interest seems to be the best indicator of a career interest in STEM after high school graduation. The authors reported a greater difficulty in attracting female students to STEM fields during high school, which means in effect that the girls had no chance to become interested by attending the STEM lessons. The early interaction with STEM fields is crucial and female students may have fewer opportunities for engagement with them or feel less welcome in these classes.

A qualitative content analysis conducted by Hewner and Knobelsdorf (2008) examined 271 biographies by Germany and US university students in CS majors and non-computing fields. Their findings detected considerable references to stereotypes in all groups, for example, identification or distinction from certain groups or stereotypes. The authors stressed the importance of identity as a motivating factor, stating "When individuals categorize themselves as members of a particular group, their view of themselves becomes dependent on their perception of the group as a whole." The authors conclude that, to have only one particular vision of the CS does not encourage self-efficacy. The more choice involved with an identity, the more likely it is to feel a sense of-belonging.

The research of Master et al. (2016) once again sees the negative stereotype as a possible reason for girls' low interest. Girls are less likely than boys to enroll in the CS courses, which are necessary to learn the basics, such as introductory CS. Interest is an important motivational variable because it can affect learning and performance outcomes and, by extension, self-confidence. The two conducted experiments (one with school girls and one with female university students), investigated whether stereotypes continue to exert influence on the interests of female students on STEM fields. The authors concluded that, e.g., stereotyped CS classrooms intensify the feeling of not belonging, which affects the girls' interest negatively. The authors then related the girls' lower sense of belonging to the perceived "Lack of Fit" or "the Sense of Not Fitting" with the stereotypes associated with CS.

### 3.2.2 Low self-confidence level during CS courses.

A survey of more than 166 undergraduate CS concentrator students at Harvard conducted by WiCS Advocacy Council (2015) shows, among other things, differences in self-confidence or self-perception towards students' programming skills. Male CS concentrators rated their confidence in programming with 3.3 on the scale of 5 after 0-6 months of experience compared to 2.6 for female CS concentrators. Only after 8 years of programming, women stated the same confidence level as men after 0-1 years of programming experience.

Direct encouragement is often missing in schools and most girls who take CS courses have already experienced some form of encouragement. A study that illustrated students' interests at the Carnegie Mellon University towards CS from 1999 to 2012 observed the following main interests across the genders (Frieze & Quesenberry, 2015): problem solving, building or creating something, working with useful applications, and long-term opportunities. These answers are from participants who are already at university level whereas teenagers mostly put CS on the same level with programming. Furthermore, the female students in this study revealed the importance of working in groups when "times get hard through project work" - this plays an important role in community-building as well. The female students at Carnegie Mellon University in general: enjoy what they do, believe it makes them feel special, and some note that to be a woman in CS is not an issue while others express their concerns that there were too few women. However, their study showed further that male students have a consistently higher confidence level concerning their coding skills than their female colleagues but explain this phenomenon also with the general tendency that women downplay their abilities. For instance, this attitude is illustrated by the statement "I feel like everyone I know performs better than I do." More than 50% of the women agreed whereas only 30% of the men did so. However, the authors found evidence of improvements of the female students' performance during the degree program. In addition, this study shows that students at Carnegie Mellon University describe a typical CS student in more positive terms, e.g., as creative, passionate, focused, smart, and diverse.

The recent study by Aivaloglou and Hermans (2019) found a significant correlation in Scratch programming courses between previous programming experiences and extrinsic motivation, CS career orientation, and self-efficacy in girls. Those who already had programming experience had an advantage in this course setting. In this study, self-efficacy and CS career orientation showed a stronger correlation in female students. Possible performance deficits in early programming courses could lead young girls to reject a CS career prematurely. This could strengthen their belief that CS is not a good fit for them. The authors conclude that students' intrinsic and extrinsic motivation and previous programming experience are important factors towards a pursuing career in CS.

The study by Alvarado et al. (2017) presents the results of a large-scale survey of students' experiences in CS classes. They observed that female students refused to ask questions in class or interacted with the instructor (felt less comfortable doing so), and as a result, they were left with lower confidence in their abilities and with doubts about the learning content. This was more the case in advanced classes with more challenging content; the authors found evidence that women prefer to ask the tutor for assistance (after the class), while men asked their male peers first. The female students also indicated their low confidence in some of the answers. For example, they disagreed more often than men with statements like, "the class pace is too slow", or "I want to be a tutor on my own next year". Nevertheless, the lower confidence was not manifested in a poorer learning outcome. The possible reason for this discrepancy could be that the female students felt that the male students were much more advanced in CS and felt too embarrassed to ask for help in front of their peers.

### 3.2.3 Differences in self-efficacy, engagement, and fun-level.

An empirical study in Los Angeles public high schools by De, Estrella, and Margolis (2006) investigated students' decisions (over 200 interviews) to study CS as a major and evaluated both the structural and psychological factors. Results showed that the factors discouraging women from CS courses included a) the image mobilized by the CS subject itself, i.e., the lonely, introvert programmer/hacker, b) that their familial or social networks are not technology-oriented, and they do not understand computing work at all, and c) they do not see the connection of how CS could support their academic and career plans. Especially the last point is critical: most found programming interesting but the decision not to pursue CS was more a strategic decision for them, e.g., to fulfill the high school graduation requirements rather than play around with computers. Furthermore, the study showed that female students are more likely to enroll in programming courses when they already know somebody who is interested or has worked in the computer field. Once again, this study shows the importance of support from home to have a positive attitude to CS beforehand, e.g., through tinkering, reading books about the topic, learning through friends or relatives, or first attempts to create their own programs. In education and especially in the context of the "Making", tinkering refers to trying things technology (including coding) out without the fear of failure, in a playful way, e.g., using Scratch or Pocket Code (Harris et al., 2016; Slany et al., 2018).

During a qualitative study among female teenagers led by Weibert et al. (2012) a more practical curriculum had been implemented. The study showed the following as determining factors for career choices in the field of CS personal attachment: sympathy for the teacher, interests of their peers, and self-concept of their CS skills. The hindering factors included badly structured CS classes and negative career aspirations. The authors developed a curriculum, in which they integrated four modules: "Sensitization and Motivation", "Product Development in Theory", "Product Development in Practice", and finally "Evaluation". Results pointed out that to link the subject of CS directly with professional IT jobs positively influences girls' interest in CS fields. In some CS courses, low engagement levels are the default norm. These courses focus not on problem-solving skills or creative skills but on basic input and output processes. Thus, the CS classes mainly rely on given examples in textbooks, with sections of code, and following directions without using logical reasoning. This setting is not only less challenging for all students, but also boring, and it even prevents them from truly understanding a programming language and the concept of coding itself. This approach does not show the multiple ways the students can use in solving the problem, does not foster any form of group work or class discussions, and finally, leads to negative experiences (for all genders). The setting that the authors of the study introduced proved helpful in positively influencing girls' interest and engagement levels. The authors conclude that framing CS as something that helps make the world a better place by creating programs and products or new ideas strongly contrasts with many of the existing negative stereotypes. The authors argue that many teenagers never quite understand what CS is and how it relates to algorithmic thinking or problem-solving. CS, for instance, is not only programming but also a tool for solving problems or creating new ideas through meaningful assignments and resources. The question female students may ask themselves is "How likely is it that I need CS knowledge in my future?" This decision influences their motivation and persistence within that academic track.

A study by Giannakos et al. (2014) consisted of three workshop programs which examined the effect of enjoyment, happiness, and anxiety in 12-year-old girls during creative development activities. The findings show that happiness had a positive effect, anxiety had a negative effect, and enjoyment had a neutral effect on students' intentions to participate in coding activities in the future. Happiness could be increased, e.g., by using humor, fun, and positive feedback, and anxiety could be decreased by praising girls' development skills and using collaborative learning environments. Female students seem to have more negative attitudes towards STEM classes, thus there is a much higher possibility to negatively rate this expectancy value.

For women, creativity and interest in STEM professions are often related, but they perceive low practical relevance of many of these subjects. A European-wide study by Microsoft (2017), in which 11,500 young women between 11 and 30 were interviewed, showed that girls between the ages of 12 and 16 are the most creative. Thus, female teenagers at this age should be supported especially in skill training and STEM. Approximately every third woman (33%) criticized how scientific topics were explained in schools, especially the topics in CS. This European research study states that creative teenage girls are particularly interested in STEM subjects and could be attracted to programming with more innovative initiatives. It is therefore important to present STEM professions as creative. Research by Khaleel et al. (2015) argues that students often find coding activities in schools difficult or boring and end up memorizing the processes without understanding them. An enjoyable approach must be adopted in learning, especially for difficult subjects. Game elements and an additional fun factor influence the general outcome of the course and make it an interesting experience for all students.

In addition, the findings by Kallia and Sentance (2018) supported the argument that boys feel significantly more efficacious in CS than girls and girls tend to underestimate their performance in comparison with boys. For their study, 123 students from seven different UK schools participated (98 male, 25 female) with the goal of completing programming tasks. The results were statistically significant in showing differences between girls' and boys' scores in self-evaluation, and differences in self-efficacy in CS and the calibration accuracy. However, the authors found no significant difference in students' performance between girls and boys.

### 3.3   Inequity in CS-Education

Gender competence, gender-specific interest guidance, or gender-sensitive and aware education are relatively new territories for many teachers and schools (Giltemeister & Robert, 2008). Especially in education, gender is constantly negotiated, produced, and reproduced by students as well as by teachers (West & Zimmermann, 1987; Basow, 2004). Teachers play an important role in the cycle of career assumptions. Gender-sensitive teachers recognize gender-stereotyped influences on students and counteract against them, reflecting also on their own teaching practices with the aim of creating equal opportunities for girls and boys.

This section summarizes articles that were grouped under RQ 3. It reviews the importance of framing supportive classrooms, starting with the teachers' roles, language used in classrooms, teaching materials, and ways of effecting change toward creating a supportive classroom environment. It concludes with an account of gender differences during coding classes, e.g., in tinkering and programming behavior.

#### 3.3.1   Teachers in CS classrooms.

A survey has been conducted by Funke et al. (2015) with 63 CS teachers in Germany asking about their experiences regarding differences between female and male students. A large number of the teachers who participated do not note any differences between genders. Some of them named differences they noted were described as the children's personalities or their working methods. For example, girls showed a more structured approach, lacked curiosity in programming, and were less self-confident about their abilities. The important question is this; did these answers already reflect their own stereotypes about how boys and girls behave in CS?

Cheryan et al. (2009) conducted different studies to prove that gender differences in interest occur when college women are confronted with stereotypical CS classrooms. First, they examined if environments can affect female students' interests. To that end, they placed stereotypical objects (e.g., science fiction posters, electronic parts) in one room and non-stereotypically associated objects (e.g., art posters, general interest books) in another. Thus primed, women exposed to stereotypical objects said that they were less interested in CS than the control group. The difference in environment had no impact on men. A stereotypical classroom increased the women's concerns about negative stereotypes, which decreasesed their sense of belonging and interest in CS courses. Even when groups consisting only of women were present in the stereotyped classroom, women's interest in the objects distributed in the environment affected whether they joined the group or not. Second, the authors tested students' decisions when applying for generic web design companies, which included these kinds of objects in their presentations. The results were the same, as both men and women identified the stereotypical environments as masculine-coded.

Concerning teachers, a study by Schwartz (2013) indicates that male students are praised more often and thought to be rebuked more often than females. On the one hand, it is important to appreciate the students' achievements, but more importantly, to formulate the feedback in such a way that it does not entail harmful attributions or tensions that would mobilize fear of failure. On the other hand, it is not helpful to praise girls for achievements that are not addressed also in boys. Praise for "normal" performance can damage self-confidence. While boys are often praised for their talents, girls are more often praised for being hard working rather than for simply being good. Overall, it is important to praise the work of female students at least in the same way as that of the male students and to provide recognition of their work done.

A study by Wong and Kemp (2017) conducted 32 semi-structured interviews with digitally skilled teenagers (aged 13–19) and observed technology career assumptions and prejudices. The researchers suggest to draw more attention to the social elements of computing; beyond the transmission of expected technical skills, teachers should reduce gender-stereotyped views (e.g., perceptions of CS as being too difficult or point to CS as a masculinized environment), and finally, teachers should promote careers in technology that are more driven by creative thinking and design (for instance, computer animation, game design, and web design). The authors point out that when students work in pairs, girls-only pairings might positively stimulate the identities of girls in computing classes through collaborative learning and peer support (but this should not be forced by the teacher). CS teachers should not strive to train the next generation of computer scientists, programmers, or technology entrepreneurs but to promote digital literacy and excellence for digital creative purposes.

The above mentioned study by Microsoft (2017) concludes that teachers' guidance and advice are major influences over students' educational incentives and attitudes. The study included 11,500 European women and found that more than half (57%) said that they had a teacher who encouraged them to pursue a STEM career. It is important for teachers to question their own stereotypes. Teachers who are not aware of inclusive and gender-sensitive education can create a classroom in which male students are placed in the more knowledgeable position by default, rather than merit.

### 3.3.2 Gender-awareness in language and CS content.

A gender-sensitive language is particularly important in languages that employ grammatical gender extensively and use the generic masculine to subsume also other genders. The feminine form is banished into a footnote, at best (Braun, 2008). The use of a gender-inclusive language is essential because

1) "Language creates pictures" and that picture is often of a masculine gender (masculine bias; by Formanowicz et al., 2015). For example, the statement "[a] scientist in his laboratory is not a mere technician: He is also a child confronting natural phenomena that impress him as though they were fairy tales" invokes the image of a male scientist in his laboratory, but it is actually a quotation from Marie Curie Sklodowska from 1937.
2) The article by Vervecken and Hannover (2015) showed that if stereotypically male occupations are spoken about using both genders instead of the generic masculine, children assess women and men equally in them (while if the terms of reference are masculine, the children rate men as more successful) and girls tend to be more interested in these professions. If companies/universities want to appeal to female talents, but are using masculine gender in language or images, they may present their programs as less attractive to women.

The study by Zagami et al. (2015) highlighted 12 different strategies to encourage female students to study CS in the US. The authors criticize that the construction and evaluation of CS programs is insufficient, stating that "[c]urrent school IT curricula do not inspire or interest girls and this is a major reason why girls do not go on to study IT further". The authors then mention the stereotypes widely used in textbooks and teaching content, and admonish the fact that no social learning in groups occurs. They call for the identification of appropriate resources and pedagogical approaches (e.g., they refer to the different brain development of the sexes) to affect change.

The research of Medel and Pournaghshband (2017) go further and examine three distinct problems in CS teaching materials to show that CS teaching materials may not be as supportive of female students. They argue that the materials do not support diverse interests or address activities preferred by women. Instead, they draw on established trends of male-centered representation, imagery, and language that may promote gender inequality. The authors proposed replacing the characters/names with animals (e.g., replace character Eve the "eavesdropper" with an owl who "watches") and employ gender-equitable imagery. The concept means using more positive images of women in examples or textbooks (e.g., instead of the objectifying imagery of "Lena" used as an example for graphic design, use standardized images, e.g., of known monuments). In language, females are often less positively portrayed and the authors suggest to use the singular pronoun "they", in the case of the English language, to refer to nouns with unspecified gender. Finally, the authors tested their more gender-equitable material in a computer security class, and their results indicated an improvement in female students' confidence in understanding the material, and no negative effect on male students.

### 3.4 Differences in coding and tinkering behavior.

Finally, this section summarizes articles that were grouped under RQ 4. Only little research exists on differences in programming between boys and girls. The study of Beckwith et al. (2006) focused on differences in feature discovery and showed that male users adopt higher amounts and different types of features compared to female users. Female students were significantly slower in trying out new features and were less likely to use them again, whereas males tried less familiar features early on. Fewer uses of features correlated with a low self-efficacy, particularly in girls. The study also refers to differences in tinkering behavior and states that male students seem to benefit more from it. However, tinkering also helps females to gain valuable information about the features and increase their self-efficacy. Low tinkering interactions and low self-efficacy occur in girls if they use environments that are described as too complex. The study concludes that gender differences exist in the way students solve problems, which may indicate a need for supportive feature designs.

Grigoreanu et al. (2008) who state that boys relate more often to testing and debugging activities also reflect this in the research. The authors conclude that there exists a gender gap in software environments, for example, in end-user programming tools (programs designed to accept user-written components in appropriate places; visual programming is one technique of end-user programming[§§§]). For instance, the authors refer to differences in problem-solving strategies; strategies used more often by women, e.g., code inspection or specification checking, are not supported sufficiently by the tool, compared to the most often used strategies by men (testing and dataflow). This affects female end-user programmers' self-efficacy, attitudes, usage of testing and debugging features, and performance. Instead of providing female users with more manuals or tutorials for the features preferred by males (i.e., forcing them to adopt the male-perspective in debugging), the authors see a greater chance for future work in supporting those features more often used by females. The study concludes that it is possible to design features that lower the barriers to female effectiveness and help to close the gender gap (e.g., with video/text strategy explanation snippets).

Furthermore, a project conducted by Zaidi, Freihofer, and Townsend (2017) exposed fifth-graders to programming. Students used the coding environment Scratch with a team of students with proficient female role models. Survey results showed that students say that computing would be equally easy and equally difficult for both boys and girls.

In addition, the study by Craig et al. (2013) observed that the participation of girls in computer classes is not the same as those of boys: Girls tend to spend more time on visual customization while boys spent more time on solving logical puzzles, and the authors point out that it is essential to consider gender differences in logical and computational skills. According to this study, it may thus be more effective to get girls interested in technology by asking them to design games rather than to focus on the learning of specific programming skills. Presentations in CS classrooms should integrate appealing visuals and interactive elements, while also having a story to tell that is fun, informative, and engaging.

One of the previously mentioned strategies by Zagami et al. (2015), points out that female students sometimes find it difficult to engage equally in traditionally male-dominated disciplines like CS. Courses which focus on girls only take advantage of the preferences for being uncompetitive and social learning opportunities can enhance female participation. On the one hand, the authors describe the positive influence of creating situations where females are preferred to males; on the other hand, this can lead again to a range of negative impacts (stereotypes, threats, discrimination, etc.). Overall, the key benefits of girls-only initiatives are social encouragement (reinforcement of CS by peers, mentors, role models), self-perception (interest in problem solving, creative and collaborative environments), and career perception (job clarity, personal relevance). Furthermore, facilitators and teachers report the difficulty in engaging girls and boys equally in traditionally male-dominated subjects such as computing. Thus, coding initiatives for girls may improve women's participation in such activities. They provide a boost in computing self-efficacy, can change stereotypes about CS, increase women's sense of belonging to the field and overall provide them with knowledge in CS. In such environments girls are not compared to boys, they are not teased, and they can communicate with each other and build their own community. If competition and hierarchy has such an impact and does not stimulate the girls to perform better but rather intimidates them, it is a good reason to separate boys and girls or try to remove the inappropriate influence. This allows collaboration and different ways of creativity. However, no progress is without its challenges. On the one hand, if computing activities for girls only are promoted in schools, first, teachers will be faced with the conflict of having to provide similar activities for boys as well, and second, it may foster the belief that girls need a special form of tutoring to learn about CS. Many authors point out that there are many reasons why we should focus on girls in CS, but lack of skills are not one of them. On the other hand, if girls see that they can do well in a girls-only environment, they also have a different standing in the mixed groups and can catch up on what boys may already have experienced earlier. The literature reveals an unequal playing field from the start as boys often have prior experience in technology, since they are more influenced from an early age (Carter, 2006). If girls can have this experience too, they can start with a healthy sense of self-confidence resulting from skill and knowledge in a previously unknown subject.

The research of Krieger, Allan, and Rawn (2015) observed tinkering strategies across genders in undergraduate students of CS via interviews and a questionnaire. According to the authors, tinkering means exploring and is generally considered as an informal practice. Thereby, they see it on the same level as using problem-solving abilities or students asking for help. Results showed different definitions or perceptions of tinkering activities by gender, and that girls are less likely to see themselves as tinkers. Thus, the authors proposed to think of teaching tinkering for non-tinkerers as well.

Miller and Webb (2015) analyzed game completion by gender and how it correlates with self-efficacy. A number of 48 "Frogger games" were created by middle school students. At the end, 60% of the students produced fully functional games, including 67% of the female students. Those who did not finish their games were less likely to see themselves as good computer problem solvers and were less likely to see themselves pursuing computer classes in the future. It is important to highlight that girls need a sense of achievement to decide to seek further computing activities.

The review of McLean and Harlow (2017) presented data from three workshops to identify the affordances of activity design that engage girls in play. They propose that in order to engage girls in coding, social relevance, storytelling, and design tasks should be part of the coding activities. Such activities provide the opportunities to participate in problem-solving by providing learners with resources and a given problem. Teachers who only focus on computer instruction may discourage many students (of all genders) from active participation.

## 4 DISCUSSION

This article reviewed possible reasons behind the low number of women in CS programs by focusing on the research of the past 14 years on female teenagers and CS education. Negative stereotypes in CS, a perceived lack of support, a sense of not belonging or not fitting in CS, a low computer self-efficacy and experience, and a lack of female role models have shown that they reduce the representation of women in CS. Therefore, the results section concentrated on articles with empirically derived reasons that provided summaries to the research questions defined in Section 2.1. When female teenagers between 12 to 15 years decide their future careers, many influences steer them away from a CS career as summarized in the image of *Figure 1*. The review confirmed that stereotypes, preconceptions, and inequality in CS classes continue to have an impact on girls. A CS classroom is often not a place that encourages females to get involved. While *Figure 1* presented the negative influences on motivation (interest, sense-of belonging, self-efficacy, and engagement/fun), Figure 2 provides a summary for a more gender conscious CS classroom by suggesting inclusive CS environments and the promotion of more balanced attributes suitable for all teenagers in CS, as it emerged from the literature that we discussed so far. Both figures can provide a useful reference for educators and researchers in the area of gender-sensitive or conscious education.

The findings of this article helped us to be more gender-conscious and to consider the gender dimension in research and CS projects at schools. Examples of such projects are summarized in Section 6.

## 5 CONCLUSION

A girl in computing today may fall into one of the two stereotypes (see *Figure 1*): A computer freak or unfit for computing. It is still true that young women who decide to enter the CS "pipeline" are pioneers or token women who have to struggle against prejudices. However, this field should not resign itself to being a discipline for men only. Women already earn about half of all university degrees, but this number includes physiology or sociology, or in STEM (Science, Technology, Engineering, and Math) disciplines, biomedical engineering or architecture, which are all already more female-dominated disciplines (National Science Foundation, 2016). The real problem exists in CS, where the percentage of female graduates has slipped every year. There is no reason that CS is reserved for men only. A better understanding of what keeps women from CS at an early age is a first step towards improving the situation.

## 6 PROJECTS INFLUENCED BY THIS REVIEW

The findings of this article influenced a project with the name "RemoteMentor" in which all the authors of this paper were involved (Spieler et al., 2020). As discussed in this article, to strengthen girls' confidence and interest, mentoring programs and appropriate role models are two key elements to introduce girls to technical subjects and to awaken their interest in CS (Archard, 2012; Stoeger et al. 2013; Beyer, 2016). Having women as mentors for other women has a huge potential, as it combines the ideas of mentoring and exposing girls to female role models. For the project "RemoteMentor", we combined both strategies in the form of online mentoring that included pairing college students with schoolgirls. The project was a one-year investigation that started in January 2018 with funding from NetIdee, a private internet foundation[****]. In this project, female students between 14 to 15 years used our app Pocket Code[††††] (an app developed at the Graz University of Technology/TU Graz) during their CS and arts lessons. Pocket Code is an Android-based visual programming language environment built to allow the creation of games, stories, animations, and many types of other apps directly on phones or tablets (Slany, 2014). The visual "lego-style" programming language used is very similar to that which is used in the previously mentioned web environment Scratch and should support users in their first programming experience. During the project, students created their ideas by using storyboards (a framework that supports them in their idea creation by defining a genre, a main character, and the interaction level with the objects). During the arts classes, students additionally chose famous paintings and created interactive memes (i.e., altered them in a creative or humorous way) through animations and games. During these regular school lessons with Pocket Code, students received real-time online mentoring by students from the TU Graz. Thus, students were able to call a mentor during a 30-40 minute session and share their screen with them to show their scripts and ask for advice (see Figure 3). The results showed that this concept is a promising approach to supporting and motivating at least a certain segment of female students. The quality of their tutoring experience depended on the combination of the teenage girls' attitudes toward programming (active or passive) and the different tutoring orientations and styles of the tutors (instructive or supportive) which led to either positive or negative outcomes of the learning experience and success for girls in playful coding experiments. One end of the spectrum occupied girls with a passive attitude to programming, the other girls with a positive, active attitude. The former were likely to benefit far less from the learning process if they were paired with tutors who merely gave instructions, instead of collaboratively supporting their mentees, than the active girls paired with a mentor expressing a supportive tutoring style (Spieler et al., 2020).

As described above, girls-only interventions were also designed on the basis of this review (Spieler, Krnjic & Slany, 2019). The intrinsic motivators identified as important for girls were integrated into a gender-sensitive framework, the Playing, Engagement, Creativity, and Creating (PECC) Framework (Spieler, 2018). The framework proposes inclusive activities during different stages of a coding activity of preparation, introduction, design, main learning, and closing; considers the gender dimension in different motivators proposed by this article; and shows how all students can benefit equally from them. To apply and evaluate the framework, in summer 2018 and 2019 one of the authors of this paper introduced the "Girls Coding Week", during which female teenagers between 12 to 15 years old used the Pocket Code app for creating gamified apps. In the first year, 13 girls took part and in the second year 28. Following the findings of the review presented here, the goal of the first day was to provide an understanding of coding and CS in general and the profession of people in CS. During this warm-up phase, girls were introduced to coding with CS-unplugged activities (i.e., to teach coding without the use of a computer, for instance, to program a classmate like a robot, [Brackmann et al., 2017]), and had discussions in groups about their coding experiences, role models, and other CS relevant topics. During the second and third days, the understanding of coding and CT-skills was delivered through ten hands-on coding units. Each unit started with a CS-unplugged activity[‡‡‡‡], a short theoretical part, a coding example created together with the whole group, and finally a coding challenge as individual work. The girls had two more activities on the fourth day: First, they were introduced to LEGO®-Mindstorms NXT robots, and second, they were able to stitch their creative programmed patterns with an embroidery machine on bags. All coding units provided the necessary preparation for the girls to create their own games on the fourth and fifth days of the workshop. The girls started with a storyboard (textual as well as graphical). First, they presented their game ideas to their peers. Second, they started creating their own games by including their own artwork. At the end of the workshop, all of the girls created their own individual games, which all included several levels with difficulties, interactivity with objects (e.g., by using different sensors of the app), and variables for points and levels. They presented their finished games in front of their parents. In both weeks we conducted questionnaires that included measures of the various factors regarding students' motivators. The quantitative evaluation shows that the values of many predictors for intrinsic motivations are located over the average. Girls agreed or strongly agreed that the coding week fostered their interests, helped them learn something about coding, they gained better knowledge of technical professions and coding, and they felt engaged while having fun during the course. Impressions of the workshop are part of Figure 4.

Furthermore, a new project "Code'n'Stitch" started in September 2018[§§§§], funded by the Austrian Research Promotion Agency (FFG/FEMtech) which introduces a gender-sensitive pedagogical framework for handicraft lessons in four Austrian secondary schools by using mobile visual pattern design with Pocket Code. Gender differences in playing behavior and game preferences raise concerns about possible gender inequalities when games are used as a motivation to explore coding. In contrast, if pupils learn to create their own patterns or geometric and artistic textile designs, they not only learn how to code but can also show the results of their code, i.e., patterns on their shirts and bag. During this project we extended the app with the option to program embroidery machines. The target group of teenage girls was involved at a very early stage of the development cycle of the app and materials. Following this research a special emphasis is given on a gender-equitable conception to consider different motivators, requirements, needs and interests of our target group.

Girls-only initiatives around the world such as GirlsWhoCode and Black Girls Code[*****] or similar initiatives had a great influence on young girls' motivation towards CS. For example, Marcu et al. (2010) conducted a CS and engineering course for middle school girls and their conclusion reflects and reinforces our findings: Girls entered the course largely with negative views on computers (e.g. "boring" and "not for me") and they finished the course with enthusiasm and appreciation for these topics and future careers options. Based on our experiences with girls in CS and by referring to this research, we plan to establish a "Girls CS Lab" during summer to perform long-term studies in order to observe any effects in the future. For this lab, we will evaluate existing offers for girls and statistics at our university and look at best practice examples (e.g., Carnegie Mellon University) with the goal to generate a united lab that follows a uniform and constructive concept. The authors hope that such a lab will support the inclusion of girls in a currently male-dominated field and will empower them to study CS, effecting a positive change in their futures.

## 7   ACKNOWLEDGEMENTS

This work has been partially funded by NetIdee Internet Foundation Austria, Netidee Projekt 2335 "RemoteMentor".

14	Spieler, Oates-Indruchovà, & SlanyBarker, L. J. & Aspray, W. (2006). The state of research on girls and IT. In Cohoon, J.M. & Aspray, W (Eds.). *Women and Information Technology: Research on underrepresentation,* Cambridge, Massachusetts institute of Technology Press, 3–54.

Basow, S. (2004). The Hidden Curriculum: Gender in the Classroom. In Michelle A. Paludi (Eds.). *Prager Guide to the Psychology of Gender* (p. 117–131). Praeger Publishers/Greenwood Publishing Group.

Beckwith, L., Burnett, M., Grigorenu, V., & Wiedenbeck, S. (2006). Gender HCI: What about the software? *IEEE Computer Society Press*, 39, 97–101, https://doi.org/10.1109/MC.2006.382

Beekhuyzen, J. & Dorries, R. (2008). *Tech Girls are Chic ! (not Just Geek!)*. tech2morrow, Brisbane.

Bell, D. A. & White, S. S. (2014). *Gender Diversity in Silicon Valley: A Comparison of Silicon Valley Public Companies and Large Public Companies*. Fenwick & West LLP.

Bers, M., Flannery, L., Kazakoff, E., & Sullivan, A. (2014). Computational thinking and tinkering: Exploration of an early childhood robotics curriculum. *Computers and Education* 72, 145–157. https://doi.org/10.1016/j.compedu.2013.10.020

Beyer, S. (2016). *Women in CS: Deterrents*. Encyclopedia of Computer Science and Technology Edition: 2ndChapter: Women in CS: Deterrents Publisher: CRC Group Editors: P. A. Laplante. https://doi.org/10.1081/E-ECST2-120054030

Bitkom (2019). Mit 10 Jahren haben die meisten Kinder ein eigenes Smartphone [At the age of 10, most children have their own smartphone]. Retrieved March 25, 2020, from https://www.bitkom.org/Presse/Presseinformation/Mit-10-Jahren-haben-die-meisten-Kinder-ein-eigenes-Smartphone

Box, S. (2018). *Digital transformation must focus on women and girls.* Policy Options Politiques, Retrieved December 20, 2018, from http://policyoptions.irpp.org/magazines/november-2018/digital-transformation-must-focus-women-girls/

Brackmann, C. P., Romàn-Gonzàlez, M., Robles, G., Moren-Leòn, J., Casali, A., & Barone, D. (2017). Development of Computational Thinking Skills through Unplugged Activities in Primary School. *Proceedings of the 12th Workshop on Primary and Secondary Computing Education*, Erik Barendsen and Peter Hubwieser (Eds.), 65–72. https://doi.org/10.1145/3137065.3137069

Braun, F. (2008) *Mehr Frauen in der Sprache. Leitfaden zur geschlechteren Formulierung* [More women in the language. Guide to gendered formulation]. Ministry of Justice, Women, Youth and Family of Schleswig Holstein.

Brewer, K. (2017). *How the tech industry wrote women out of history.* The Guardian. Retrieved October 20, 2018, from https://www.theguardian.com/careers/2017/aug/10/how-the-tech-industry-wrote-women-out-of-history

Buskens, I. & Webb, A. (2014). *Women and ICT in Africa and the Middle East: Changing
Selves, Changing Societies*. In Zed Books.

Carter, L. (2006). Why students with an apparent aptitude for computer science don't choose to major in computer science. *Proceedings of the 37th SIGCSE technical symposium on Computer science education*, 27–31. https://doi.org/10.1145/1121341.1121352

Charleston, L. J. (2017). *Ways The Tech Industry In Australia Is Trying To Attract And Retain Females*. Huffpost. Retrieved September 04, 2019, from https://www.huffingtonpost.com.au/2017/03/31/ways-the-tech-industry-in-australia-is-trying-to-attract-and-ret_a_22019419/

Cheryan, S., Plaut, V. C., Handron, C., & Hudso, L. (2013). The stereotypical computer scientist: Gendered media representations as a barrier to inclusion for women. *Springer Science & Business Media,* 69(1-2), 58–71. https://doi.org/10.1007/s11199-013-0296-x

Cheryan, S., Siy, J. O., Vichayapai, M., Drury, B. J., & Kim, S. (2011). Do female and male role models who embody STEM stereotypes hinder women's anticipated success in STEM? *Social Psychological and Personality Science,* 2(6), 656–664. https://doi.org/10.1177/1948550611405218

Cheryan, S., Plaut, V., Davies, P., & Steele, C. (2009). Ambient belonging: How stereotypical cues impact gender participation in computer science. *Journal of Personality and Social Psychology,* 97(6), 1045–1060. https://doi.org/10.1037/a0016239

Ciobanu, C. (2018). *Eastern Europe, heaven for women in tech*. NewsMavens, Central & Eastern Europe. Retrieved September 20, 2018, from https://newsmavens.com/news/aha-moments/1326/eastern-europe-heaven-for-women-in-tech.

Clarke-Midura, J., Allan, V., & Close, K. (2016). Investigating the Role of Being a Mentor as a Way of Increasing Interest in CS. *Proceedings of the 47th ACM Technical Symposium on Computing Science Education*, 297–302. https://doi.org/10.1145/2839509.2844581

Cohoon, J. M., & Aspray, W. (2006). *Women and information technology: Research on underrepresentation*, The MIT Press.

Committee on European Computing Education (2017). *Informatics Education in Europe: Are We All In The Same Boat?* Report by The Committee on European Computing Education. Jointly established by Informatics Europe & ACM Europe. Retrieved September 20, 2019, from https://www.informatics-europe.org/component/phocadownload/category/10-reports.html?download=60:cece-report

Craig, A., Coldwell-Neilson, J., & Beekhuyzen, J. (2013). Are IT interventions for Girls a Special Case? *Proceeding of the 44th ACM technical symposium on Computer science education*, 451–456. https://doi.org/10.1145/2445196.2445328

Dalberg (2015). *Empowering Women to Use the Mobile Internet*, Dalberg Design, Retrieved September 20, 2018, from https://dalberg.com/our-ideas/empowering-women-use-mobile-internet

Dasgupta, N. & Stout, J. G. (2014). Girls and Women in Science, Technology, Engineering, and Mathematics. *Policy Insights from the Behavioral and Brain Sciences* 1(1), 21–29. https://doi.org/10.1177/2372732214549471

Dele-Ajayi, O., Emembolu, I., Peers, M., Shimwell, J., & Strachan, B. (2018). Exploring Digital Careers, Stereotypes and Diversity with Young People through Game Design and Implementation. *In 2018 IEEE Global Engineering Education Conference (EDUCON) IEEE*. https://doi.org/10.1109/EDUCON.2018.8363301

De, J., Estrella, R., & Margolis, J. (2006). Lost in translation: Gender and high school computer science. *In Aspray, W. & Cohoon, J.M. (Eds.)* Women in IT: Reasons on the Reasons of Under-Representation. MIT Press.

Engler, S. & Wieland H. F. (1995). Ent-Dramatisierung der Differenzen. Studentinnen und Studenten der Technikwissenschaften [De-dramatization of differences. Students of the engineering sciences]. Bielefeld: Kleine Verlag.
14

## 9  Positionality

**Spieler** has a PhD in Engineering Sciences. She has been a visiting professor at the Institute for Mathematics and Applied Informatics at the University of Hildesheim (Germany), as well as the head of the Department of Computer Science Didactics, since November 2019. Previously, she worked at Graz University of Technology (Austria) and she is still part of the Product Owner Board of the TU Graz Catrobat project. Her path into CS was far from conventional. Spieler did not have her first programming courses until she reached the university level, where many of her colleagues were already advanced programmers. Supportive professors and classmates were important in her success, but she also credits a willful obliviousness to her unusual development. She believes that many teenage girls struggle with the question of whether they fit into the study of CS. Her goal is to enhance girls' first experiences and to spark creativity with an experiential and discovery learning approach, which allows for independent thinking and new ways of constructing information. By exploring new concepts and standards in educational technology and CS education, she aims to promote learning throughout the life span and help everyone, especially students and teachers, to cope in the new digital age. https://bernadette-spieler.com

**Oates-Indruchova** is a full professor of sociology of gender. She grew up under state socialism, a political regime that pushed aggressively for technocratic and (to a substantial degree) gender-equal education: the core subjects in her academic-oriented high school were mathematics and physics, followed by biology and chemistry, while critical thinking and, particularly, any social sciences were heavily restricted. She has undergraduate background in both humanities (English) and life sciences (sports studies) and became acquainted with the early word processors in the early 1980s. As soon as the political climate in her home country changed in 1990, she began to pursue her humanities and, later, social science interests as a graduate student. She is also a trained high school teacher by original profession and taught for a decade in a teacher-training college, before she moved to a more research-oriented career. Nevertheless, the early education and also her interest in languages very much provided a prism for the RemoteMentor project: programming is a language and learning a language in a school setting is a familiar context for her.

**Slany** is passionate about poverty alleviation through coding education for teenagers, in particular girls, refugees, and teenagers in developing countries, directly on their personal mobile phones. Slany's Catrobat non profit free open source project develops educational smartphone apps since 2010. The apps work in a sustainable way even for teenagers in less privileged areas who do not have access to PCs and laptops, by relying on the phones that most teenagers around the world already own personally. This bypasses traditional school pedagogy and instead uses a constructivist approach that focuses on developing game applications and fun. Professionally, Slany is a full professor and is consulting, teaching, and doing research on sustainable large scale agile software development and user experience topics for mobile platform projects.

## 10  Figures

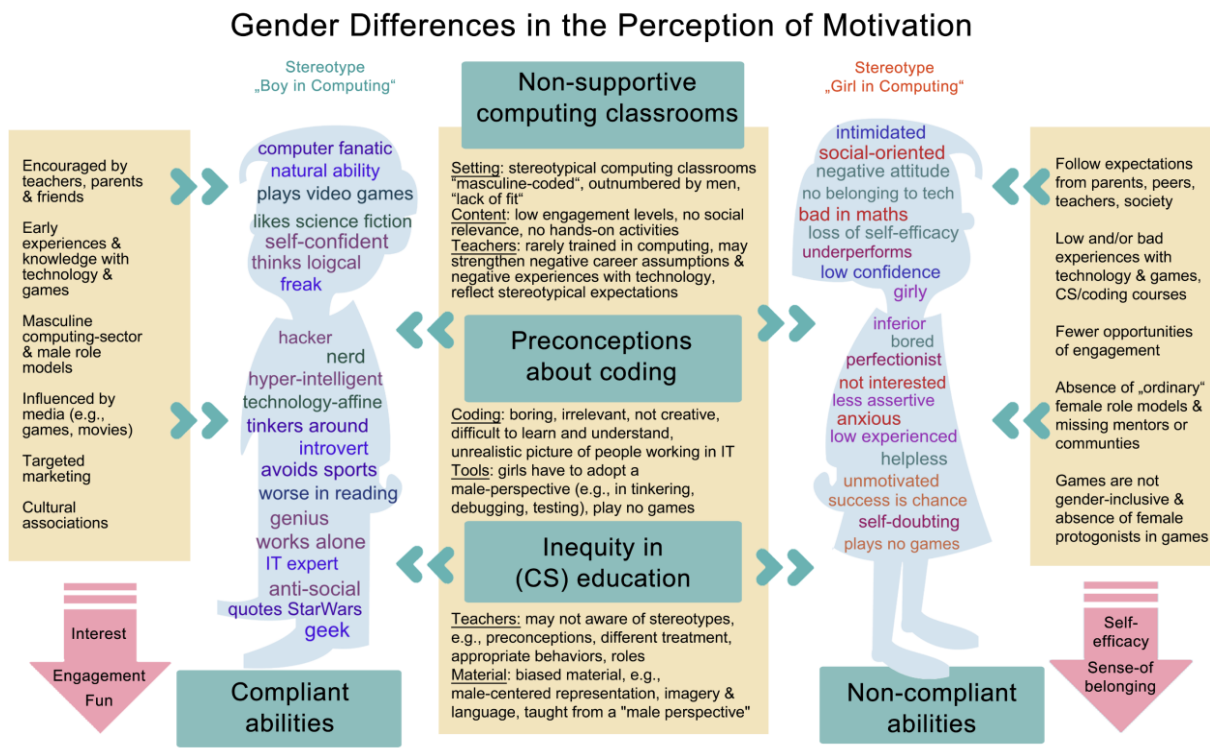

**Figure 1**: The decrease in motivation in girls: Social and cultural shaping of the negative stereotype of a "Girl in Computer Science" by its binary relation to the stereotyped "Boy in Computer Science" by and through the preconceptions and inequities in (CS-) education. Consequently, the motivators (interest, sense-of belonging, engagement/fun, and self-efficacy) are reduced in children and teenagers who act against that norm. The articles used for this review represent published scientific literature on the topics of gender stereotypes in computing and on differences in coding classes at high school and university levels that shape the stereotypes of boys and girls in CS along with the attributes associated with both stereotypes.
Sources:
Cultural and social influences: (Jenson, de Castell & Fisher, 2007; Cheryan et al., 2011; Cheryan et al., 2013; Young et al., 2013; Dasgupta & Stout, 2014; Galdi et al., 2014; Gabay-Egozi et al., 2015; Frieze & Quesenberry, 2015; Lynch et al., 2016; Master et al., 2016; Lewis, Anderson & Yasuhara, 2016; Semmens, Piech & Fried, 2016)
Preconceptions and differences in (CS-) classrooms: (Carter, 2006; De, J., Estrella, R. & Margolis, 2006; Weibert et al., 2012; Giannakos et al., 2014; Alvarado et al., 2017; Aivaloglou & Hermans, 2019)
Inequity in CS-education: (Grigoreanu et al. 2008; Cheryan et al., 2009; Funke et al., 2015; Zagami et al., 2015; Medel & Pournaghshband, 2017; Dele-Ajayi et al., 2018)
Internet sources: (Schwartz, 2013; Brewer, 2017; Microsoft, 2017; Zoo, 2017; Google, 2018,)

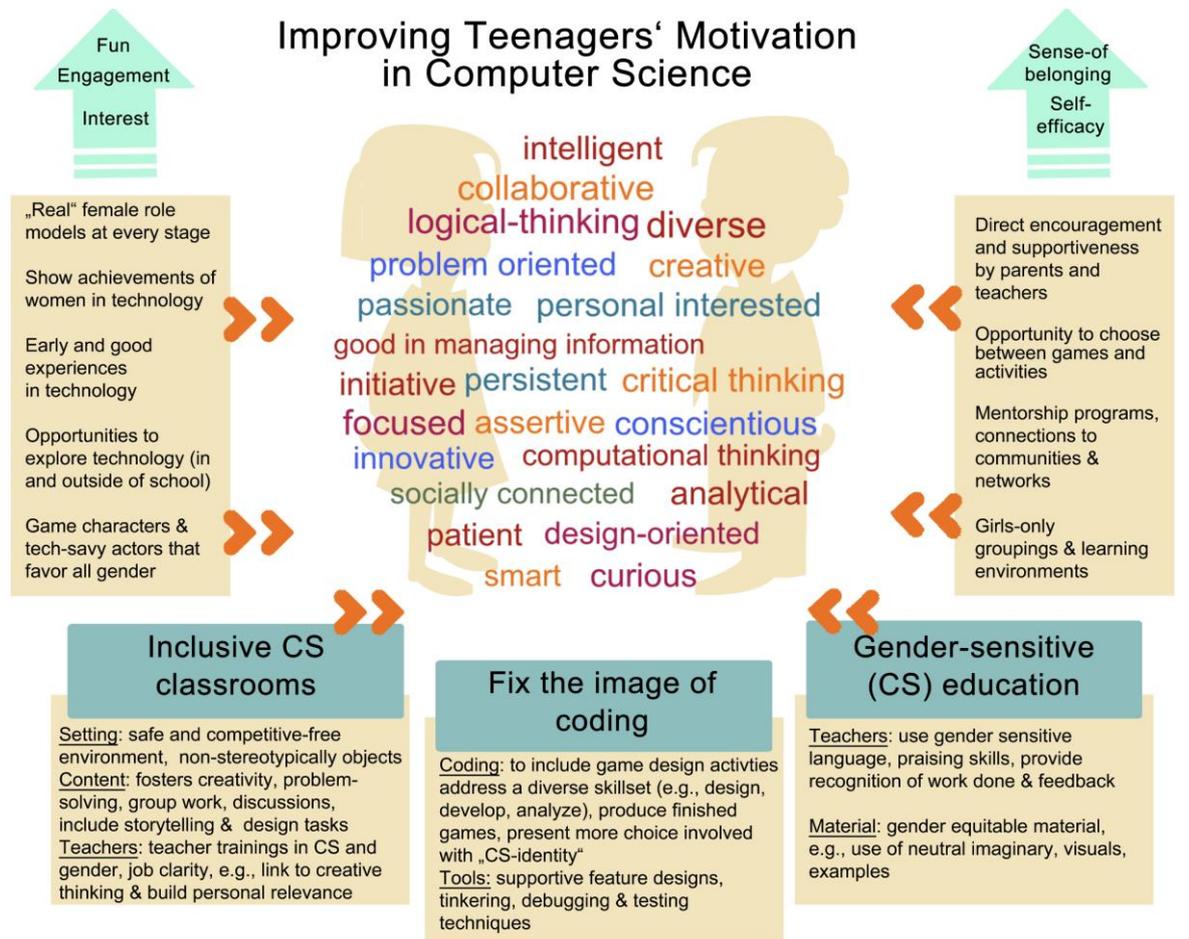

**Figure 2**: The increase in motivation by creating CS classes free from stereotypes: reshaping the picture of people in computer science and building non-stereotypical classrooms by gender-aware teachers.
Sources:
Cultural and social influences: (Lockwood, 2006; Jenson, de Castell & Fisher, 2007; Cherian et al., 2013; Semmens, Piech & Fried, 2016; Young et al., 2013, 2017)
Preconceptions and differences in (CS-) classrooms: (Carter, 2006; Sadler et al., 2012; Weibert et al., 2012; Giannakos et al., 2014; Frieze & Quesenberry, 2015; Khaleel et al., 2015; Unfried, 2015)
Inequity in CS-education: (Beckwith et al., 2006; Grigoreanu et al., 2008; Craig et al., 2013; Krieger, Allan & Rawn, 2015; Zagami et al., 2015; Alvarado et al., 2017; McLean & Harlow, 2017; Wong & Kemp, 2017)
Internet sources: (Schwartz, 2013; Microsoft, 2017; Yee, 2017; Google 2018)

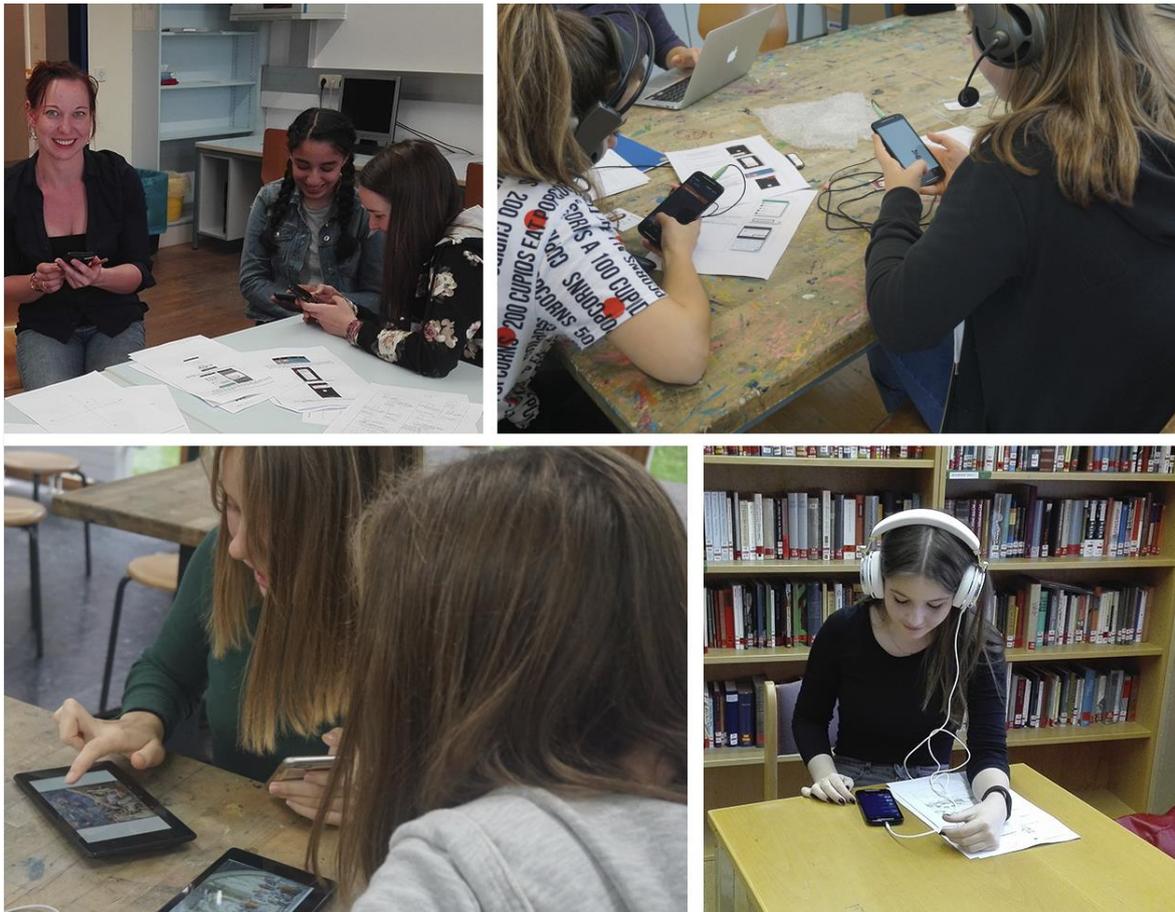

**Figure 3:** "RemoteMentor" project funded by NetIdee. Female students get remote mentoring by university students through screen-sharing and audio transmission during their arts lesson.

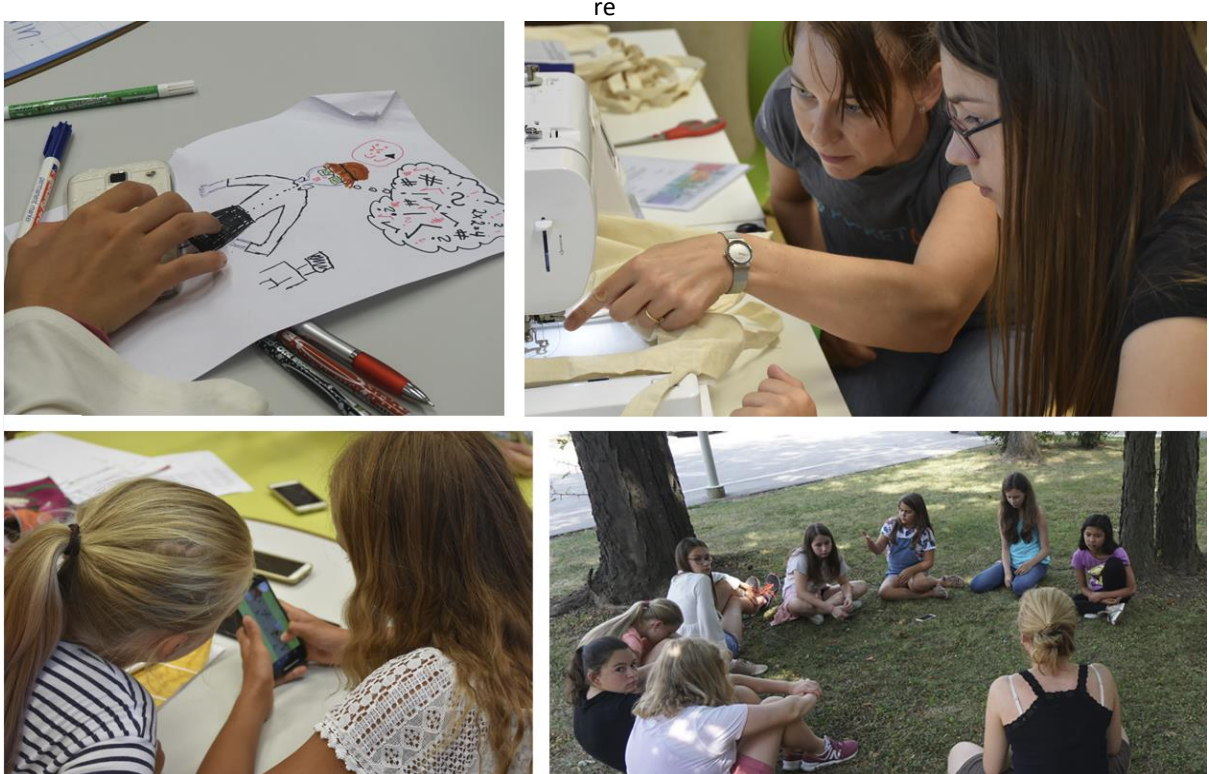

**Figure 4**: Girls Coding Week 2018. a) talking about different professions in computing to get rid of stereotypes , b) use "robots" like a programmable embroidery machine, c) Game creation with our app Pocket Code (https://catrobat.at/pc), and d) unplugged coding activities.

# 11  Tables

**Table 1:** Definition of the general concepts by using the PICOC method (Petticrew & Roberts, 2006)

| | |
|---|---|
| Population | Students between 12 to 15 years old<br>Students in CS major at university |
| Intervention | Girls in CS, Girls in coding and game design<br>Female students CS degree programs/coding |
| Comparison | Boys and girls |
| Outcomes | Differences in perception of motivation |
| Context | Academic, school and off- context |

**Table 2:** Complete List of Selected Studies ACM (19), PsycNET (2), Wiley (1), Springer (2), IEEE Xplore (4)

| No. | Author(s) | Title | Year | Database | RQ |
|---|---|---|---|---|---|
| 1 | Aivaloglou & Hermans | Early Programming Education and Career Orientation: The Effects of Gender, Self-Efficacy, Motivation and Stereotype | 2019 | IEEE | 4 |
| 2 | Alvarado & Minnes | Gender Differences in Students' Behaviors in CS Classes throughout the CS Major | 2017 | ACM | 2, 3 |
| 3 | Beckwith et al. | Gender HCI: What about the software? | 2006 | IEEE | 3 |
| 4 | Carter | Why students with an apparent aptitude for computer science don't choose to major in computer science | 2006 | ACM | 2 |
| 5 | Cheryan at al. | The stereotypical computer scientist: gendered media representations as a barrier to inclusion for women | 2013 | Springer | 1 |
| 6 | Cheryan, et al. | Ambient belonging: How stereotypical cues impact gender participation in computer science | 2009 | PsycNET | 1, 3 |
| 7 | Clarke-Midura, Allan & Close | Investigating the Role of Being a Mentor as a Way of Increasing Interest in CS | 2016 | ACM | 1 |
| 8 | Craig, Coldwell-Neilson & Beekhuyzen | Are IT interventions for Girls a Special Case? | 2013 | ACM | 3 |
| 9 | Funke, Berges, Mühling & Hubwieser | Gender differences in programming: research results and teachers' perception | 2015 | ACM | 3 |
| 10 | Giannakos, Jaccherie & Leftheriotis | Happy Girls Engaging with Technology: Assessing Emotions and Engagement Related to Programming Activities | 2014 | Springer | 2 |
| 11 | Grigoreanu et al. | Can feature design reduce the gender gap in end-user software development environments? | 2008 | IEEE | 3 |
| 12 | Hewer & Knobelsdorf | Understanding Computing Stereotypes with Self-Categorization Theory | 2008 | ACM | 1 |
| 13 | Jenson, de Castell & Fisher | Girls Playing Games: Rethinking Stereotypes | 2007 | ACM | 1, 4 |
| 14 | Khaleel et al. | The Study of Gamification Application Architecture for Programming Language Course | 2015 | ACM | 2 |
| 15 | Krieger, Allen & Rown | Are Females Disinclined to Tinker in Computer Science? | 2015 | ACM | 3 |
| 16 | Ko & Davis | Computing Mentorship in a Software Boomtown: Relationships to Adolescent Interest and Beliefs | 2017 | ACM | 1 |
| 17 | Kallia & Sentance | Are boys more confident than girls? The role of calibration and students' self-eff icacy in programming tasks and computer science | 2018 | ACM | 2 |
| 18 | Lewis, Anderson & Yasuhara | "I Don't Code All Day": Fitting in Computer Science When the Stereotypes Don't Fit | 2016 | ACM | 1 |
| 19 | Lynch et al. | Strong, and Secondary: A Content Analysis of Female Characters in Video Games | 2016 | Wiley | 3 |
| 20 | Master, Sapna & Meltzoff | Computing Whether She Belongs: Stereotypes Undermine Girls' Interest and Sense of Belonging in Computer Science | 2016 | PsycNET | 2 |
| 21 | McLean & Harlow | Designing Inclusive STEM Activities: A Comparison of Playful Interactive Experiences Across Gender | 2017 | ACM | 3 |
| 22 | Medel & Pournaghshband | Eliminating Gender Bias in Computer Science Education Materials | 2017 | ACM | 3 |
| 23 | Miller and Webb | Game Design: Whose game works at the end of the day? | 2015 | ACM | 4 |
| 24 | Semmens, Piech & Fried | Who Are You? We Really Wanna Know... Especially If You Think You're Like a Computer Scientist | 2016 | ACM | 1 |
| 25 | Stanko & Zhirosh | Young women who choose IT: what role do their families play? | 2017 | IEEE | 1 |
| 26 | Weibert, von Rekowski & Festl | Accessing IT: a curricular approach for girls. In Proceedings of the 7th Nordic Conference on Human-Computer Interaction: Making Sense Through Design | 2012 | ACM | 2 |

| 27 | Zaidi, Freihofer & Townsend | Using Scratch and Female Role Models while Storytelling Improves Fifth-Grade Students' Attitudes toward Computing | 2017 | ACM | 4 |
|---|---|---|---|---|---|
| 28 | Zagami et al. | Girls and computing: Female participation in computing in schools | 2015 | ACM | 2 |

---

\* Brebas Challenge Austria: http://wettbewerb.biber.ocg.at/, Germany: https://bwinf.de/biber/, UK: http://www.bebras.uk/).
† Code.org: https://hourofcode.com/ and Hour of Code: https://code.org/diversity
‡ GirlsWhoCode: https://www.codefirstgirls.org.uk
§ Code:First Girls: https://www.codefirstgirls.org.uk/)
\*\* Scratch programming environment: https://scratch.mit.edu
†† Examples for mentoring programs: http://computermentors.org/kidscode/, coder dojo: https://www.coderdojoparramatta.org/, https://locations.sylvanlearning.com/us/mentor-oh/coding-for-kids
‡‡ Tomb Raider: https://www.tombraider.com
§§ Horizon Zero Dawn: https://www.guerrilla-games.com/play/horizon
\*\*\* Sword and Sorcery: https://play.google.com/store/apps/details?id=com.capybaragames.sworcery
††† Grand Theft Auto: https://dotesports.com/the-op/news/grand-theft-auto-vi-news-21158
‡‡‡ Lollipop Chainsaw: http://lollipopchainsaw.com/
§§§ End-user programming: http://www.cs.uml.edu/~hgoodell/EndUser/whatsEUP.htm
\*\*\*\* NetIdee project Remote Mentor: https://www.netidee.at/remotementor
†††† Pocket Code app: https://catrob.at/pc
‡‡‡‡ CS unplugged: https://csunplugged.org
§§§§ Code'n'Stitch project: https://catrob.at/codeNstitch
\*\*\*\*\* Black Girls Code: http://www.blackgirlscode.com/